\documentstyle[12pt]{article}
\def\hybrid{\topmargin 0pt      \oddsidemargin 0pt
        \headheight 0pt \headsep 0pt
        \textheight 9in         % US paper
        \textwidth 6.25in       % A4 paper
        \marginparwidth .875in
        \parskip 5pt plus 1pt   \jot = 1.5ex}

\catcode`\@=11
\def\marginnote#1{}
\newcount\hour
\newcount\minute
\newtoks\amorpm
\hour=\time\divide\hour by60
\minute=\time{\multiply\hour by60 \global\advance\minute by-\hour}
\edef\standardtime{{\ifnum\hour<12 \global\amorpm={am}%
        \else\global\amorpm={pm}\advance\hour by-12 \fi
        \ifnum\hour=0 \hour=12 \fi
        \number\hour:\ifnum\minute<10 0\fi\number\minute\the\amorpm}}
\edef\militarytime{\number\hour:\ifnum\minute<10 0\fi\number\minute}
\def\draftlabel#1{{\@bsphack\if@filesw {\let\thepage\relax
   \xdef\@gtempa{\write\@auxout{\string
      \newlabel{#1}{{\@currentlabel}{\thepage}}}}}\@gtempa
   \if@nobreak \ifvmode\nobreak\fi\fi\fi\@esphack}
        \gdef\@eqnlabel{#1}}
\def\@eqnlabel{}
\def\@vacuum{}
\def\draftmarginnote#1{\marginpar{\raggedright\scriptsize\tt#1}}
\def\draft{\oddsidemargin -.5truein
        \def\@oddfoot{\sl preliminary draft \hfil
        \rm\thepage\hfil\sl\today\quad\militarytime}
        \let\@evenfoot\@oddfoot \overfullrule 3pt
        \let\label=\draftlabel
        \let\marginnote=\draftmarginnote
   \def\@eqnnum{(\theequation)\rlap{\kern\marginparsep\tt\@eqnlabel}%
\global\let\@eqnlabel\@vacuum}  }
\def\numberbysection{\@addtoreset{equation}{section}
        \def\theequation{\thesection.\arabic{equation}}}
\def\underline#1{\relax\ifmmode\@@underline#1\else
        $\@@underline{\hbox{#1}}$\relax\fi}
\def\titlepage{\@restonecolfalse\if@twocolumn\@restonecoltrue\onecolumn
     \else \newpage \fi \thispagestyle{empty}\c@page\z@
        \def\thefootnote{\fnsymbol{footnote}} }
\def\endtitlepage{\if@restonecol\twocolumn \else  \fi
        \def\thefootnote{\arabic{footnote}}
        \setcounter{footnote}{0}}  %\c@footnote\z@ }
\catcode`@=12
\relax

\def\beq{\begin{equation}}
\def\eeq{\end{equation}}
\def\bea{\begin{eqnarray}}
\def\eea{\end{eqnarray}}
\def\nn{\nonumber}

\relax
\hybrid
\begin{document}
\begin{titlepage}
\begin{center}
February~1995 \hfill    PAR--LPTHE 95/05 \\[.4in]
{\large\bf RENORMALIZATION GROUP SOLUTION FOR\nn\\ THE TWO-DIMENSIONAL
RANDOM BOND POTTS MODEL WITH BROKEN REPLICA SYMMETRY}\\[.3in]
	{\bf Vik.S.Dotsenko}\\
	{\it Landau Institute for Theoretical Physics,\\
	Russian Academy of Sciences,\\
        Kosygina 2, 117940  Moscow, Russia}\\
        {\bf Vl.S.Dotsenko\footnote{Also at the Landau Institute for
	Theoretical Physics, Moscow}, M. Picco and
	P.Pujol} \\
	{\it LPTHE\/}\footnote{Laboratoire associ\'e No. 280 au CNRS}\\
       \it  Universit\'e Pierre et Marie Curie, PARIS VI\\
       \it Universit\'e Denis Diderot, PARIS VII\\
	Boite 126, Tour 16, 1$^{\it er}$ \'etage \\
	4 place Jussieu\\
	F-75252 Paris CEDEX 05, FRANCE\\
\end{center}
\vskip .1in
\centerline{\bf ABSTRACT}
\begin{quotation}
We find a new solution of the renormalization group for the Potts model
with ferromagnetic random valued coupling constants. The solution exhibits
universality and broken replica symmetry. It is argued that the model
reaches this universality class if the replica symmetry is broken
initially.  Otherwise the model stays with the replica symmetric
renormalization group flow and reaches the fixed point which has been
considered before.
\end{quotation}
\end{titlepage}
\newpage

The problem of new critical behavior induced by randomness in spin systems
has a considerable history. Starting with a classical $\varphi^{4}$
problem, the modified critical behavior has later been studied for the
two-dimensional Ising and Potts models by various renormalization group
techniques, and by numerical simulations. Incomplete list of references is
provided in [1-12]. Replicas has been used generally to deal with the
quenched disorder and replica symmetric solutions have generally been
looked for. The first example of replica symmetry broken solutions of the
renormalization group has been suggested in [13], in the context of the
$\varphi^{4}$ model.  In this letter we report on the replica symmetry
broken solution for the two-dimensional Potts model with random bonds. The
model reaches this solution if the replica symmetry is broken
initially. In
contrast, the two-dimensional Ising model turns out to be stable with
respect to replica symmetry breaking \cite{dotb2}. It reaches always the
replica symmetric critical behavior which has been studied earlier [5-12].

For theoretical study one uses models with a weak disorder, e.g. models
with spin couplings having small fluctuations around a mean ferromagnetic
value. This gives a possibility to study the model in continuum, because
one reaches the critical point sufficiently close before the randomness
becomes important. For the two-dimensional Potts model in particular this
allows to use eventually the renormalization group based on the conformal
theory of the unperturbed model. In this approach the effective theory
could be described by the Hamiltonian
\beq
H = H_{0} +\int d^{2}x\,
m(x)\varepsilon(x)
\eeq
where $H_{0}$ represents, symbolically, the
conformal theory of the unperturbed model, while the second term with a
spatially random mass $m(x)$ coupled to the energy operator represents the
effective randomness due to spatially inhomogeneous coupling constants of
spins. Replicating the model and taking the average of the partition
function over $ m(x) $ one gets the effective homogeneous theory with the
Hamiltonian:
\beq H = \sum_{a=1}^{n}H_{0}^{a} + g\int d^{2}x\sum_{a\neq
b}\varepsilon_{a}(x) \varepsilon_{b}(x)
\eeq
where $ g $ is defined by
\beq
\langle m(x)m(x')\rangle = g\delta(x-x')
\eeq
Without loss of generality
one could assume a Gaussian distribution for $ m(x) $ because the terms in
the effective Hamiltonian produced by higher moments are irrelevant, in the
sense of the renormalization group. Also, a spatially uncorrelated
distribution of coupling constants, and so of $ m(x) $, is being
assumed. For extra details on the definition of this model the reader could
consult the papers [8,12].

The $\beta$ - function of the renormalization group for the model (2) has
been derived, up to second order of perturbation theory (third order in
$g$), in [8], with the following result:
\beq
\frac{dg}{d\xi} = \beta(g)
\eeq
\beq
\beta (g) = -3\epsilon g + 4\pi(n-2)g^{2}-16\pi^{2}(n-2)g^{3}+O(g^{4})
\eeq
Here $\xi$ is the renormalization group log-scale parameter, and $\epsilon$
is related to the central charge of the conformal theory for the Potts
model at the critical point, on a homogeneous lattice. The following
parametrization is being used [12]:
\beq
c = 1 - 24\alpha_{0}^{2} = 1 - 6(\alpha_{+} + \alpha_{-})^{2}
\eeq
\beq
\alpha_{\pm}=\alpha_{0}\pm\sqrt{\alpha_{0}^{2}+1},
\quad\alpha_{+}\alpha_{-}=-1
\eeq
\beq
\alpha_{+}^{2}=\frac{4}{3}+\epsilon
\eeq
The case of $\alpha_{+}^{2}=\frac{4}{3},\, c=\frac{1}{2}$ corresponds
to the conformal theory for the Ising model. By shifting $\alpha_{+}^{2}$
one shifts the central charge $c$. In particular, for the 3 - components
Potts model, the parameters of the associated conformal theory are~:
\beq
c = \frac{4}{5},\quad\alpha_{+}^{2}=\frac{6}{5}
\eeq
which corresponds to
\beq
\epsilon = -\frac{2}{15}
\eeq
The energy operator $\varepsilon$ of the Potts model corresponds, in the
conformal theory classification, to the operator $\Phi_{1,2}$[14].
In general, the dimensions of the conformal operators $\Phi_{n',n}$ are
given by the Kac formula [15]:
\beq
\Delta_{n',n}=\alpha_{n',n}^{2}-2\alpha_{0}\alpha_{n',n}=\frac{(\alpha_{-}
n' + \alpha_{+}n)^{2}-(\alpha_{+}+\alpha_{-})^{2}}{4}
\eeq
\beq
\alpha_{n',n}=\frac{1-n'}{2}\alpha_{-}+\frac{1-n}{2}\alpha_{+}
\eeq
For $\varepsilon \sim\Phi_{1,2}$ one gets
\beq
\alpha_{1,2}=-\frac{\alpha_{+}}{2}
\eeq
\beq
\Delta_{\varepsilon}=2\Delta_{1,2}=2
(\alpha_{1,2}^{2}-2\alpha_{0}\alpha_{1,2})=\frac{3}{2}\alpha_{+}^{2}
-1=1+\frac{3}{2}\epsilon
\eeq
So, for the case of the Ising model, $\epsilon =0,\,\Delta_{\varepsilon}
=1$, the perturbation in (2) is marginal, and then one defines the
renormalization group for the Potts model in terms of the $\epsilon $ -
expansion technique [8].

The renormalization of the operator $\varepsilon (x)$ has also been found
in [8], up to the second order, with the following result:
\beq
\varepsilon(x)\rightarrow\varepsilon_{ren}(x)=
Z_{\varepsilon}(\xi)\varepsilon
(x)
\eeq
\beq
\frac{d\log Z_{\varepsilon}}{d\xi}=\gamma_{\varepsilon}(g)
\eeq
\beq
\gamma_{\varepsilon}(g)=4\pi(n-1)g-8\pi^{2}(n-1)g^{2}+O(g^{3})
\eeq
 Finally, the renormalization of the spin operator has been found in
[12], up to the third order:
\beq
\sigma(x)\rightarrow\sigma_{ren}(x)=Z_{\sigma}(\xi)\sigma(x)
\eeq
\beq
\frac{d\log Z_{\sigma}}{d\xi}=\gamma_{\sigma}(g)
\eeq
\bea
\gamma_{\sigma}(g)=-3(n-1)\pi^{2}\epsilon\left(1+
2\frac{\Gamma^{2}(-\frac{2}{3})
\Gamma^{2}(\frac{1}{6})}{\Gamma^{2}(-\frac{1}{3})\Gamma^{2}
(-\frac{1}{6})}\right) g^{2}\nn\\+4(n-1)(n-2)\pi^{3}g^{3}+O(g^4)
\eea
Here $\Gamma(z)$ is the Euler $\Gamma$-function.

Using results for $\beta,\gamma_{\varepsilon} $ and $\gamma_{\sigma}$,
eqs. (5),(17),(20), where one puts eventually $n=0$, the following
results have been obtained for $g_{c},\,\Delta_{\varepsilon}$ [8] and
$\Delta_{\sigma} $
[12]:
\beq
g_{c}=-\frac{3}{8\pi}\epsilon + \frac{9}{16\pi}\epsilon^{2} +
O(\epsilon^{3})
\eeq
\beq
\Delta'_{\varepsilon}=\Delta_{\varepsilon}-\gamma_{\varepsilon}(g_{c})=
\Delta_{\varepsilon}-\frac{3}{2}\epsilon + \frac{9}{8}\epsilon^{2} + O
(\epsilon^{3})=1 + \frac{9}{8}\epsilon^{2} + O(\epsilon^{3})
\eeq
\beq
\Delta'_{\sigma}=\Delta_{\sigma} - \gamma_{\sigma}(g_{c})=\Delta_{\sigma}
-\frac{27}{32}\frac{\Gamma^{2}(-\frac{2}{3})\Gamma^{2}(\frac{1}{6})}
{\Gamma^{2}(-\frac{1}{3})\Gamma^{2}(-\frac{1}{6})}\epsilon^{3}
+O(\epsilon^{4})
\eeq
For the 3-component Potts model, $\epsilon=-\frac{2}{15}$, one gets the
following numerical values:
\beq
\Delta'_{\epsilon}=1,02+O(\epsilon ^3)
\eeq
\beq
\Delta'_{\sigma}=\frac{2}{15}+0,00132+O(\epsilon ^4)
=0,13465+O(\epsilon ^4)
\eeq

In the solution just outlined one assumes that the replica symmetry is not
broken initially, and then it is preserved by the renormalization
group. Physical arguments for a more general approach, namely to start with
a Hamiltonian in which the replica symmetry is lifted, have been suggested
in [13]. For the case of the Potts model this amounts to replacing the
effective Hamiltonian (2) with:
\beq
H=\sum_{a=1}^{n}H_{0}^{a}+\int d^{2}x\sum_{a\neq b}g_{ab}\varepsilon_{a}(x)
\varepsilon_{b}(x)
\eeq
where $g_{ab}$ is a Parisi type matrix [16]. We shall assume next this
conjecture and give the solution of the theory (26), in order to allow,
on the basis of the results obtained, to verify the conjecture itself.

By using the techniques of [12] the generalization of the renormalization
group equations is straightforward. Eqs. (4) and (5) are replaced by
\beq
\frac{dg_{ab}}{d\xi}=\beta_{ab}
\eeq
\beq
\beta_{ab}=-3\epsilon g_{ab}+4\pi(g^{2})_{ab}-16\pi^{2}((g^{2})_{aa}g_{ab}
-(g_{ab})^{3})
\eeq
Here $(g^{2})_{ab}=\sum_{c}g_{ac}g_{cb}$. The fixed point matrix $g_{ab}$
should satisfy the equation:
\beq
-3\epsilon g_{ab} + 4\pi(g^{2})_{ab}-16\pi^{2}((g^{2})_{aa}g_{ab}-
(g_{ab})^{3})=0
\eeq
We always assume the diagonal elements of the coupling matrix $g_{ab}$
to be zero. This is because the corresponding terms could always be
absorbed  into $\sum_{a=1}^{n}H_{0}^{a}$, in eq.(26). Then, for the
Parisi type  matrices, one has in general the following rules [16]:
\beq
g_{ab}\rightarrow g(x)
\eeq
\beq
(g^{2})_{ab}\rightarrow-2g(x)\int_{0}^{1}dyg(y)
-\int_{0}^{x}dy(g(x)-g(y))^{2}
\eeq
\beq
(g^{2})_{aa}\rightarrow-\int_{0}^{1}dy  g^{2}(y)
\eeq
Here $ x $ is the continuous parameter which replaces the matrix indices:
$g_{ab}\sim g(a-b)\sim g(x) $. Putting $\tau = 3|\epsilon|$ ($\epsilon $
is assumed to be negative in its definition (8)), replacing $ g\rightarrow
\frac{1}{4\pi}g$, and using the prescriptions (30) - (32), one gets from
(29) the following equation for $ g(x) $:
\beq
\tau g(x) - 2\bar{g}g(x) - \int_{0}^{x}dy(g(x)-g(y))^{2}+g^{3}(x)+g(x)
\bar{g^{2}}=0
\eeq
Here $\bar{g}=\int_{0}^{1}dy g(y),\, \bar{g^{2}}=\int_{0}^{1}dy g^{2}(y)$.
Note that the structure of the fixed-point equation (33) coincides with
the saddle-point equation for the Parisi order parameter function
in the infinite-range spin-glasses near the phase transition point
(the parameter $\tau$ in eq.(33) corresponds to the reduced temperature
$\tau = (1 - T/T_{c}) << 1$ in the spin-glass model)[16].
The solution of this equation is straightforward.
Taking a derivative with respect to $x$ one gets:
\bea
\tau g'(x)-2\bar{g}g'(x)-2g'(x)\int_{0}^{x}dy(g(x)-g(y))\nn\\+3g'(x)
g^{2}(x)+g'(x)\bar{g^{2}}=0
\eea
So, either
\beq
g'(x)=0
\eeq
or
\beq
\tau-2\bar{g}-2\int_{0}^{x}dy(g(x)-g(y))+3g^{2}(x)+\bar{g}^2 = 0
\eeq
Differentiating again one gets
\beq
-2g'(x)x+6g'(x)g(x)=0
\eeq
There are two solutions:
\beq
g(x)=const\equiv g_{1}
\eeq
and
\beq
g(x)=\frac{1}{3}x
\eeq
Next, we take $g(x)$ in the form:
\beq
g(x)=\cases{{1\over3}x ,& $0<x<x_{1}$\cr g_{1} ,& $x_{1}
<x<1$ \cr}
\eeq
with
\beq
x_{1}=3g_{1}
\eeq
Then, we put back $g(x)$ into
the original equation (33). In particular
\beq
\bar{g}=g_{1}-\frac{3}{2}g_{1}^{2},\quad \bar{g^{2}}=g_{1}^{2}-2g_{1}^{3}
\eeq
Substituting (40),(42) into (33), either for $0<x<x_{1}$ or for $x_{1}
<x<1$ one gets, after some simple algebra
\beq
g_{1}\approx\frac{1}{2}\tau +\frac{1}{2}\tau^{2},\quad
x_{1}\approx\frac{3}{2}\tau +\frac{3}{2}\tau^{2}
\eeq
up  to the second order in $\tau = 3|\epsilon|$.
This solution can be compared with the replica symmetry one,
\beq
g_{ab}\sim g(x)=const=\frac{1}{2}\tau +\frac{1}{4}\tau^{2},\,\,0<x<1
\eeq
given by eq.(21) (after a rescaling
$g_{ab}\rightarrow\frac{1}{4\pi}g_{ab})$.

Note that in terms of the original "dynamical" eqs.(27),(28)
the {\it continuous} fixed-point solution, eq.(40), is the only one which
is (marginally) attractive. This is guaranteed by the fact that
the eigenvalues spectrum of the corresponding Hessian of the
infinite-range spin-glass problem is known to be non-positive.
On the other hand, all the other non-trivial fixed-point solutions
which have step-like structure (they correspond to finite number
of RSB steps in the replica matrix $g_{ab}$) are unstable because
for any {\it finite} number of steps there exist positive finite
eigenvalues [16].

We can now find the dimensions of the operators $\varepsilon$ and
$\sigma$, for the solution (40). Again, with a straightforward
generalization,  one finds for $\gamma_{\varepsilon}(g)$ and
$\gamma_{\sigma}(g)$ the following expressions:
\beq
\gamma_{\varepsilon}(g)= 4 \pi {1\over n} \displaystyle \sum_{ab} g_{ab}
-8 \pi^{2} (g^{2})_{aa} + O(g^3)
\eeq
\beq
\gamma_{\sigma}(g)=3 \pi^{2}\epsilon\left(1+2\frac{\Gamma^{2}
(-\frac{2}{3}) \Gamma^{2}(\frac{1}{6})}{\Gamma^{2}(-\frac{1}{3}) \Gamma^{2}
(-\frac{1}{6})}
\right)
(g^{2})_{aa}\nn\\+8\pi^{3}(g^{3})_{aa}+O(g^4)
\eeq
Using the Parisi ansatz $ g_{ab}\rightarrow g(x) $ with the rules
(30)-(32) and, in addition,
\beq
(g^{3})_{aa}\rightarrow\int_{0}^{1}dx(xg^{3}(x)+3g(x)\int_{0}^{x}dy\,
g^{2}(y))
\eeq
one obtains the following expressions:
\beq
\gamma_{\varepsilon}(g)= -4 \pi \int\limits_{0}^{1} g(x) dx
+8 \pi^{2} \int\limits_{0}^{1} g^2(x) dx + O(g^3)
\eeq
$$
\gamma_{\sigma}(g)=-3 \pi^{2}\epsilon\left(1
+2\frac{\Gamma^{2}(-\frac{2}{3}) \Gamma^{2}(\frac{1}{6})}
{\Gamma^{2}(-\frac{1}{3})\Gamma^{2}(-\frac{1}{6})} \right)
 \int\limits_{0}^{1} g^2(x) dx
$$
\beq
+8\pi^{3} \int\limits_{0}^{1} \left( x
g^3(x)+ 3 g(x)  \int\limits_{0}^{x} g^2(y) dy \right) dx
+O(g^4)
\eeq
Simple analysis of these expressions shows that the modification of
$\Delta'_{\sigma}$ for the solution (40) for $ g(x)$ will be of order
$\epsilon^{4}$. As we haven't kept the $\sim\epsilon^{4}$ terms in
$\gamma_{\sigma}(g)$ the accuracy is not sufficient to use the modified
$g(x)$. To $\epsilon^{3}$ order, $\Delta'_{\sigma}$ remains the same,
eqs.(23),(25).
On the other hand, the accuracy is sufficient for $\gamma_{\varepsilon}(g)$
which depends of terms of order $\sim g$ and $\sim g^{2}$ with coefficients
of order $\sim 1 $. Simple calculation with
$\gamma_{\varepsilon}(g)$ in (45) leads to
\beq
\Delta''_{\varepsilon}=\Delta_{\varepsilon}-\gamma_{\varepsilon}(g)=
\Delta_{\varepsilon}-\frac{3}{2}\epsilon +O(\epsilon^{3})=1+O(
\epsilon^{3})
\eeq
which can be compared to the solution of $\Delta'_{\varepsilon}$ given by
eq.(24): $\Delta'_{\epsilon}=1,02+O(\epsilon ^3)$

To conclude, we have found an explicit form, at the order $\epsilon^{2}$,
of replica symmetry broken fixed point in the 3-states Potts model with
random  ferromagnetic bonds. We have also calculated one observable
quantity, the dimension of the energy operator $\Delta''_{\varepsilon}$,
which could distinguish this universality class in a numerical experiment
[17].
%%%%% ***************************************** %%%%%%%%

\end{document}